\documentclass[aps,prb,twocolumn,floatfix,superscriptaddress,showpacs,amsmath,amssymb]{revtex4-1}
\usepackage{graphicx}
\usepackage{epsfig}
\allowdisplaybreaks

\newcommand{\be}{\begin{equation}}
\newcommand{\ee}{\end{equation}}
\newcommand{\bea}{\begin{eqnarray}}
\newcommand{\eea}{\end{eqnarray}}
\newcommand{\ba}{\begin{eqnarray*}}
\newcommand{\ea}{\end{eqnarray*}}

\newcommand{\m}[1]{\mathcal{#1}}
\newcommand{\eps}{\varepsilon}

\begin{document}

\title{Transport Across an Impurity in One Dimensional Quantum Liquids Far From Equilibrium}
\author{Marco Schir\'o}
\affiliation{Institut de Physique Th\'{e}orique, CEA, CNRS-URA 2306, F-91191,
Gif-sur-Yvette, France}
%\affiliation{Department of Chemistry, Columbia University, New York, New York 10027, U.S.A.}
\author{Aditi Mitra}
\affiliation{Department of Physics, New York University, 4 Washington
  Place, New York, New York 10003, USA}
\begin{abstract}
We study the effect of a single impurity on the transport properties of a one dimensional quantum liquid highly excited away from its ground state
by a sudden quench of the bulk interaction. In particular we compute the time dependent dc current to leading order in the impurity potential,
using bosonization, and starting both from the limit of a uniform system, and from the limit of two decoupled semi-infinite systems.
Our results reveal that the nonequilibrium excitation of bulk modes induced by the global quench has important  effects on the conductor-insulator
quantum phase transition, turning it into a crossover
for any small quench amplitude, and destroying the exact duality between the
conducting and insulating fixed points. 
In addition, the current displays a faster decay towards the steady-state as compared to the equilibrium case,
a signature of quench-induced decoherence.
\end{abstract}
\date{\today}
%\pacs{72.10.Pm,05.70.Ln,72.15.Qm}
\maketitle

\section{Introduction}

Transport phenomena in strongly correlated quantum systems are typically sensitive to interactions, inhomogeneities and low-dimensionality and often reveal intriguing and unexpected effects. A prominent example is the impact of a single potential barrier in a one dimensional interacting electron liquid. While the clean system would display, despite interactions, ideal conductance quantization in the presence of
metallic leads~\cite{MaslovStonePRB95,SafiSchulzPRB95,PonomarenkoPRB95,RoschAndreiPRL00}, the impurity at zero temperature is capable of turning
the perfect conductor into an insulating link, giving
rise to power law corrections to transport coefficients at finite temperature or
finite voltage bias~\cite{KaneFisherPRL92,KaneFisherPRB92,MatveevYueGlazmanPRL93}. The physics of a single impurity in a Luttinger liquid has attracted enormous attention in the past decades~\cite{FendleyLudwigSaleurPRL95,FendleyLudwigSaleurPRB95,ChangRMP_ChiralLL} and
still finds beautiful applications in systems such as the edges of a quantum spin Hall
insulator~\cite{WuBernevigZhangPRL06,IlanEtAlPRL12} or in electronic quantum circuits~\cite{SafiSaleurPRL04,JezouinNatComm12,SouquetSafiSimonPRB13}.
Traditional condensed matter transport settings involve a system initially in thermal equilibrium which is perturbed away from it by the action of
some external field conjugate to a conserved current. Within linear response theory one then obtains transport coefficients,
which give information on the structure of the underlying ground-state and its low-energy excitations.

Recently, experimental advances in controlling and probing ultra-cold atomic gases has offered a new platform to study nonequilibrium time-dependent phenomena
in a fully tunable setting~\cite{Bloch_rmp08,LangenGeigerSchmiedmayerArxiv14}. First generation of experiments studied the dynamics induced by rapidly changing in time some
system parameter~\cite{Greiner_nature_02_bis,DeMarco_PRL11,TrotzkyNatPhys12} in an otherwise isolated system, a so called quantum quench.
More recently, the experimental focus shifted towards realizing genuine transport experiments with
cold atoms~\cite{BrantutEtAlScience12,KrinnerEtAlArxiv14}. It is important to realize that these systems are very well isolated from their environment and hence intrinsically out of equilibrium, therefore the standard condensed matter idealizations do not necessarily apply.

Motivated by these experimental developments, in this paper we investigate the transport properties of nonequilibrium quantum many body states which are thermally isolated and highly excited above their ground state.
While for generic, non-integrable and ergodic, quantum many-body systems one may expect the excitation energy to be effectively converted into temperature at long times~\cite{Lux14,TavoraRoschMitraPRL14}, and hence to induce decoherence, the situation is less clear at intermediate time scales where long-lived prethermal states may
emerge~\cite{Berges04,Kehrein_prl08,SchiroFabrizio_prl10,GringScience12,Karrasch_PRL12,SandriSchiroFabrizioPRB12,
MitraPRB13,TavoraRoschMitraPRL14}, especially parametrically close to an integrable point.
The idea we pursue in this work is to use transport as a probe to unveil the structure of these prethermal excited states,
to understand the relevant excitations and whether  non-trivial quantum phenomena survive at these high-energies, and if so, in which form. It should be noted that there has been a recent spur of interest towards understanding similar dynamical quantum correlations in isolated many body systems excited after
quantum quenches~\cite{FoiniCugliandoloGambassi_PRB11,EsslerEvangelistiFagottiPRL12,FoiniCugliandoloGambassi_JStatMech12,
RossiniEtAlEPL14}. We have considered an example along these lines in a recent work on a time-dependent orthogonality catastrophe
problem~\cite{SchiroMitraPRL14} where we introduced a novel dynamical Loschmidt echo, encoding the response of a highly excited state to a local perturbation. Here we study a similar problem from the point of view of transport.

Specifically, in this paper we compute the time-dependent current of a one dimensional system
of spinless electrons described by the Luttinger model, which is excited by a sudden change of the two-particle interaction and a simultaneous switching-on
of a local scattering potential. These two perturbations have very different effects, the former injecting extensive energy into the bulk modes,
the latter creating a non-linear channel for local dissipation.  We discuss the physics both in the limit of a uniform system, where the local
potential induces a weak backscattering term, and in the limit of two decoupled semi-infinite systems where we switch on a local tunneling. In both cases, 
using bosonization, we formulate the problem in terms of a boundary sine-Gordon model in a nonequilibrium transient environment. 
Using perturbative approaches we discuss the fate of the conducting-insulating zero 
temperature quantum phase transition in the presence of a nonequilibrium bulk excitation,
thus complementing and expanding previous results~\cite{PerfettoStefanucciCiniPRL10,KennesMedenPRB13}.

The results for the steady state current reveal the emergence of a novel energy scale associated with a
quench-induced decoherence effect, which turns the sharp equilibrium transition into a smooth crossover. Interestingly, the very same energy scale has
been shown to play a key role in the transient orthogonality catastrophe problem, cutting off the renormalization group flow of the backscattering and
resulting in an exponential decay of the Loschmidt echo~\cite{SchiroMitraPRL14}.

Our analysis, although perturbative in nature, ultimately suggests that the steady state impurity problem in a quenched Luttinger model has very 
different properties than in equilibrium at zero temperature. In particular the quench acts as a relevant perturbation on both sides of the equilibrium 
transition, even when the local non-linearity would be irrelevant at the tree level, generating an effective decoherence which drives the 
system away from the uniform and open-chain fixed points. While this behavior is reminiscent of an 
effective temperature, we will see that this analogy is only qualitative rather than quantitative, as the scaling of 
physical quantities in the steady state with respect to this emergent energy scale do not show signature of non-trivial power-laws like the ones encountered at 
non-zero temperatures.
At the same time we find that deviations from the asymptotic steady state regime, due to transient effects at finite time, 
display nonequilibrium power laws with characteristic Luttinger liquid exponents which may be a 
signature of a sort of Luttinger liquid universality out of equilibrium, as recently discussed~\cite{Karrasch_PRL12,KennesMedenPRB13,KennesPRL14}.

Note that the problem of an impurity in a quenched Luttinger model has been recently addressed 
in Refs.~\onlinecite{PerfettoStefanucciCiniPRL10} and \onlinecite{KennesMedenPRB13}. 
In the former, the dynamics of the current in the tunneling regime was investigated after a global quench of the bulk interaction and a faster decay to the stationary state was found. In the latter, results for the temperature scaling of the conductance in the steady state after the quench have been obtained using different approaches such as bosonization and functional RG. Although we do not address temperature dependence of transport coefficients in this paper, and we
work always at zero temperature, it is useful to comment on the relation between the results of Ref.~\onlinecite{KennesMedenPRB13} and the physical picture emerging from this work. We will do this in the discussion section.

Finally, it is worth mentioning that the question of nonequilibrium effects on local quantum criticality has been addressed recently also
in the context of driven quantum systems~\cite{RibeiroSiKirchner_EPL13} and specifically for single impurity in Luttinger liquids in the context of noisy driven Josephson junction circuits~\cite{DallaTorreetalPRB2012} or tunneling in biased
quantum wires~\cite{DinhBagretsMirlinPRB10} using the out-of-equilibrium bosonization framework~\cite{Gutman_PRB2010,NgoBagretsMirlinPRB13}.
The picture emerging in these cases, where the nonequilibrium perturbation induces a decoherence mechanism which eventually cuts the quantum critical 
power-law behavior, is consistent with our results for the isolated quenched problem.

The paper is organized as follows. In section~\ref{sect:WBS} we introduce the system, the nonequilibrium protocol,
and a derivation of an expression for the time dependent current in terms of a bosonic Green's function. We then evaluate
the current in the
weak-backscattering limit in section~\ref{sect:wbs_correction}. Such a perturbative approach eventually breaks down
in a certain region of parameter space and this will lead us in
section~\ref{sect:Tunn} to formulate the problem in the limit of two disconnected 1D systems coupled by a local tunneling term.
We will compute perturbatively the tunneling current in section~\ref{sect:tunn_correction} and thus obtain a complete picture of the problem.

\section{Transport in the Weak-Backscattering Regime}\label{sect:WBS}

We start by discussing the model and the nonequilibrium set up.
We consider a one dimensional system of interacting spinless fermions described by the Tomonaga-Luttinger (TL) model. We assume the system
to be initially (at time $t\leq 0$) in the ground-state $\vert\Psi_0\rangle$ of the TL Hamiltonian
\begin{eqnarray}
&&H_0=H_{\rm free}+\frac{g_4^0}{2}\int dx\left(\rho_L^2(x)+\rho_R^2(x)\right)\nonumber\\
&&+g_{2}^0\int dx\rho_L(x)\rho_R(x)
\end{eqnarray}
where $H_{\rm free}=v_F\sum_{\alpha=L,R}s_{\alpha}\int dx\psi^{\dagger}_{\alpha}i\partial_x\psi_{\alpha}$ with $s_{L}=-s_R=1$,
$\rho_{\alpha}=\psi^{\dagger}_{\alpha}\psi_{\alpha}$ is the fermion density with given chirality $\alpha=L/R$, and
$g^0_4(g^0_2)$ are the strengths of the intra (inter)-branch scattering processes.

At time $t=0$ we assume a sudden change (quench) in the value of the bulk interactions
to $g_2,g_4$, so that the system will evolve with the Hamiltonian
\begin{eqnarray}
&&H= H_{\rm free}+
\frac{g_4}{2}\int dx\left(\rho_L^2(x)+\rho_R^2(x)\right)\nonumber\\
&&+g_{2}\int dx\rho_L(x)\rho_R(x)
\end{eqnarray}
This global quench injects extensive energy into the system and triggers a nonequilibrium occupation of the bulk modes. Dynamics
arising simply due to this quench has been discussed extensively in the literature~\cite{Cazalilla_prl06,Cazalilla_long09,MitraGiamarchiPRL11}.
In addition to the bulk excitation, we assume that a static impurity potential $V_{\rm loc}$ has also been
suddenly switched on at time $t=0$. This static impurity can induce an intra-branch $(R,L)\rightarrow (R,L)$ as well as 
inter-branch $(R,L)\rightarrow(L,R)$ scattering of fermions, that we write as the sum of two contributions
\bea\label{eqn:Vimp}
V_{\rm loc}= V_{\rm fs}\left(\psi^{\dagger}_L(0)\psi_{L}(0)+L\leftrightarrow R\right)+\\+
V_{\rm bs}\left(\psi^{\dagger}_L(0)\psi_{R}(0)+h.c.\right)\,.
\eea
the first term representing the forward scattering while the latter the backward scattering.
As a result the wave function at time $t$ is $\vert\Psi(t)\rangle = \exp\left(-iH_{+}t\right)\vert\Psi(0)\rangle$, with $H_+=H+V_{\rm loc}$.

We employ bosonization to describe the system, thus introducing the bosonic
fields $\phi(x)$ and $\theta(x)$ describing collective density and current excitations respectively,
\bea
\psi_{L/R}(x)\sim \frac{1}{\sqrt{2\pi\alpha}}\,e^{i\left(\theta(x)\pm\phi(x)\right)}\\
\rho_{L/R}(x)=-\partial_x\left(\phi(x) \pm\theta(x)\right) /2\pi
\eea
The Hamiltonians before and after the quench in terms of these bosonic fields are
\be
H_0 = \frac{u_0}{2\pi}\int dx
\left[K_0\left(\partial_x\theta(x)\right)^2 + \frac{1}{K_0}\left(\partial_x \phi(x)\right)^2
\right]
\ee
and
\be
H_+ =  \frac{u}{2\pi}\int dx
\left[K\left(\partial_x\theta(x)\right)^2 + \frac{1}{K}\left(\partial_x \phi(x)\right)^2\right]+V_{\rm loc}
\ee
Within bosonization, the impurity potential in Eq.~(\ref{eqn:Vimp}) can be written as
\be
V_{\rm loc}=g_{\rm fs}\,\partial_x\phi(0) + g_{\rm bs}\,\cos 2\phi(0)\label{Vloc1}
\ee
where the effective impurity couplings read respectively~\cite{Giamarchi_2003}
\be
g_{\rm fs}=-\frac{V_{\rm fs}}{\pi} \qquad g_{\rm bs}=\frac{V_{\rm bs}}{\pi \alpha}
\ee
The Luttinger parameter $K_0,K$ and the sound velocities
$u_0,u$ are related to the Fermi velocity and the interaction parameters as
\bea\label{eqn:u0K0}
u_0 =v_F\,\sqrt{\left(1+g^0_4/2\pi v_F\right)^2-(g^0_2/2\pi v_F)^2}\\
K_0 = \sqrt{\frac{1+g^0_4/2\pi v_F-g^0_2/2\pi v_F}{1+g^0_4/2\pi v_F+g^0_2/2\pi v_F}}
\eea
with similar relations holding for $u,K$ as
a function of $g_2,g_4$. In order to preserve Galilean invariance we choose the values of $g_2,g_4$ such that $u_0K_0=uK$.
This amounts to simply requiring that $g_4-g_2=g_4^0-g_2^0$.

In order to probe transport through the system we will study linear response to
a weak electric field $E(x,t)$ applied
after the quench,
\be
E(x,t)=-\partial_t\,A(x,t)
\ee
with $A(x,t)$ being the vector potential. In the presence of the electric field, the LL Hamiltonian is modified according to the minimal substitution which amounts to the shift
\be
\partial_x\theta(x) \rightarrow \partial_x\theta(x) - eA(x,t)
\ee
The current operator can be obtained as the functional derivative of $H(A)$ with respect to the vector potential
\be
J(x,t)=-\frac{\delta H}{\delta A(x,t)}
\ee
As usual, the current  has two contributions, $J(x,t)=J_d(x,t)+J_p(x,t)$, the diamagnetic ($J_d$)
and the paramagnetic ($J_p$) one. The former is given by (hereafter $\hbar=1$)
\be
J_d(x,t)=-\frac{e^2 u K}{\pi} A(x,t)\equiv  -D A(x,t)
\ee
where we have introduced the diamagnetic term $D=e^2 u K/\pi $.
The latter is
\be
J_p(x,t) =\frac{e uK}{\pi}\partial_x\theta(x,t)=\frac{e}{\pi}\partial_t\phi(x,t)
\ee
Finally, we notice that the total current can be equivalently obtained from the continuity equation
$$
\partial_t\rho(x,t)+\partial_x J(x,t)=0
$$
by computing the time derivative of the density $\rho(x,t)=-(e/\pi)\partial_x\phi(x,t)$
$$
\partial_t\rho(x,t)=-\frac{ie}{\pi}[H(A),\partial_x\phi(x,t)]
$$
using the LL Hamiltonian with the minimal substitution. The result of this calculation obviously
recovers the expression for the current $J(x,t)=J_d(x,t)+J_p(x,t)$.

We can now compute the average current within linear response theory, by doing perturbation theory in $A(x,t)$.
Since the diamagnetic part of the current $J_d(x,t)$ is already linear in the vector potential, we only need to take into account the paramagnetic contribution which gives
\be\label{eqn:Jp}
\langle J_p(x,t)\rangle=-\int_0^\infty dt'\int dx'\chi^R_{JJ}(xt;x't')\,A(x',t')
\ee
where we have defined the retarded current-current correlation function
\bea\label{eqn:chi_jj}
\chi^R_{JJ}(xt;x't') =
-i\theta(t-t')\langle\,[J_p(x,t),J_p(x',t')]\rangle \nonumber\\
=-i\frac{e^2}{\pi^2}\theta(t-t')\langle\,[\partial_t\phi(x,t),\partial_{t'}\phi(x',t')]\rangle
\eea
and extended the time integral up to infinity. We stress here that the average in Eq.(\ref{eqn:chi_jj}) is taken over the initial density matrix (ground state of $H_0$) while the operators are evolved with the Hamiltonian $H_+$. The resulting correlator is not the usual equilibrium one and in particular it is not time-translational invariant. This reflects the effect of the global and local quantum quenches that have been performed
on the system at time $t=0$.
Finally we have for the current
\be\label{eqn:current}
\langle J(x,t)\rangle=-D A(x,t)+\langle J_p(x,t)\rangle
%-\int_0^{\infty}dt'\int dx'\chi^R_{JJ}(xt;x't')A(x',t')
\ee
We now use an important identity relating the current-current correlation~(\ref{eqn:chi_jj}) to the retarded Green's function of the field $\phi(x,t)$ defined as
\be
\mathcal{G}^R(xt;x't')=- i\theta(t-t')\langle\,[\phi(x,t),\phi(x',t')]\rangle
\ee
We have
\be
\chi^R_{JJ}(xt;x't') = \frac{e^2}{\pi^2}\partial_{t'}\partial_t\mathcal{G}^R(xt;x't')-2D\delta(x-x')\delta(t-t')
\ee

Substituting this result into Eq.~(\ref{eqn:current}) and using Eq.~(\ref{eqn:Jp}), we find that the last term cancels the diamagnetic term exactly, and after
an integration by parts we obtain
\be
\langle J(x,t)\rangle=-\frac{e^2}{\pi^2} \int_0^t dt'\int dx'\,\partial_t\mathcal{G}^R(xt;x't')\,E(x't')
\ee
This is in principle an exact result for the time dependent current through a finite-size LL and we can now further specify the electric field profile. We assume a sudden switching of the field, $E(x,t)=\theta(t) E(x)$ and in addition,
since we are ultimately interested in the current due to the impurity~(\ref{eqn:Vimp}), we assume that the potential
drop occurs around $x=0$, $V(x)=V\,\theta(-x)$, so that the electric field is effectively a delta-function
$E(x,t>0)=-\partial_x\,V(x)=V\delta(x)$. Then, the current at $x=0$,  $\langle J(0,t)\rangle\equiv I(t)$ can be written only in terms of the exact retarded Green's function of the local field at the impurity site,
\be
G^R(t,t')=-i\theta(t-t')\langle[\phi(t),\phi(t')]\rangle
\ee
with $\phi(t)\equiv\phi(0,t)$, and it reads
\be\label{eqn:current_wbs}
I(t)=-\frac{e^2V}{\pi^2} \int_0^t dt'\,\partial_t G^R(t,t')
\ee
This equation, which gives the time-dependent current in terms of an exact dynamical correlator of the local field, is the
main result of this section. It generalizes to the time-dependent quench problem the well known equilibrium result~\cite{Giamarchi_2003}.  From this we can compute the current for a pure LL, using the expression for the bare
retarded local Green's function
\be
G_0^{R}(t>t')=-K\arctan\Lambda\left(t-t'\right)
\ee
where $\Lambda=u/\alpha$ is an ultra-violet cut-off.
We obtain
\be
I_0(t)=\frac{e^2K V}{\pi^2}\arctan\Lambda t
\ee
which approaches in the long time limit $I_{\rm 0ss}/V=e^2K/2\pi=e^2K/h$ as expected. In the next section we will evaluate the corrections to the time dependent current due to the local potential. The forward scattering term in Eq.~(\ref{Vloc1}) turns out to not contribute to the current, as we discuss explicitly in appendix~\ref{sect:fwd_scatt}, therefore in the
next section we will focus our attention on the backward scattering term which is responsible for the interesting physical effects
with $V_{\rm loc}=g_{\rm bs} \cos(2\phi(0))$.

\subsection{Weak-Backscattering Corrections to Linear Conductance}\label{sect:wbs_correction}

We are now in the position to evaluate the weak back-scattering correction to the conductance. We just need to evaluate the local Green's function for the field $\phi(t)$
\be
G_{ab}(t,t')=-i\langle\phi_a(t)\phi_b(t')\rangle
\ee
where $a,b=\pm$ are Keldysh indices. We have defined the various Green's functions and the relations
between them in appendix~\ref{sect:GF}. To lowest order in the backscattering $g_{\rm bs}$ we obtain~\cite{TavoraAditiPRB14}
\be\label{eqn:deltaG}
\delta G_{ab}(t,t')=
-\frac{g^2_{\rm bs}}{2}
\int dt_1dt_2\sum_{cd}(cd)
F_{cd}(t_1,t_2)
%\Lambda^{cd}_{ab}(t,t';t_1,t_2)
\Lambda^{cd}_{ab}(t_1,t_2)
\ee
where we define $\delta G_{ab}=G_{ab}-G_{0,ab}$, 
$$
F_{cd}(t_1,t_2)=-i\langle \cos 2\phi_c(t_1)\cos 2\phi_d(t_2)\rangle
$$
and
\begin{widetext}
\be
\frac{\Lambda^{cd}_{ab}(t_1,t_2)}{4}
=G_{0,ac}(t,1)G_{0,cb}(1,t')+G_{0,ad}(t,2)G_{0,db}(2,t')
%+\nonumber\\
-G_{0,ac}(t,1)G_{0,db}(2,t')-G_{0,ad}(t,2)G_{0,cb}(1,t')
%&&\nonumber
\ee
\end{widetext}
The retarded component is given by
\be
G^R(t,t') =-\frac{1}{2}\sum_{ab}\,b\,G_{ab}(t,t')\,,
\ee
Substituting the resulting expression
into Eq.~(\ref{eqn:current_wbs}), and after some algebra we obtain  for the current $\delta I(t)=I(t)-I_0(t)$
\be\label{eqn:deltaIt_0}
\delta I(t)= -C
%\frac{4e^2V}{\pi^2}g_{\rm bs}^2
\int_0^t dt_1\partial_t G_0^R(t,t_1)\int_0^{t_1} dt_2
F^R(t_1,t_2)\,\Phi(t_1,t_2)
\ee
with $C=\frac{4e^2V}{\pi^2}\,g_{\rm bs}^2$ and
\bea
\Phi(t_1,t_2)&=& K\int_{t_2}^{t_1}\,d\tau\,\arctan\Lambda\tau=K\Lambda t_1\arctan\Lambda t_1+\nonumber\\
&&-K\Lambda t_2\arctan\Lambda t_2 -\frac{K}{2}\log\left(\frac{1+\Lambda^2 t_1^2}{1+\Lambda^2t_2^2}\right)
\nonumber
\eea
The retarded correlator entering the expression for the current is found to be
\be
F^R(t_1>t_2)=
-\frac{\sin\left(2K\arctan\Lambda(t_1-t_2)\right)}
{\left(1+\Lambda^2\left(t_1-t_2\right)^2\right)^{K_{\rm neq}}}
f(t_1,t_2)
\ee
where we have introduced
\be
f(t_1,t_2)=\left[
\frac{\left(1+4\Lambda^2 t_1^2\right)\left(1+4\Lambda^2 t_2^2\right)}
{\left(1+\Lambda^2(t_1+t_2)^2\right)}\right]^{K_{\rm tr}/2}
\ee
and the exponents
\be\label{eqn:KneqKtr}
K_{\rm neq}=\frac{K_0}{2}\left(1+\frac{K^2}{K_0^2}\right)\qquad
K_{\rm tr}=\frac{K_0}{2}\left(1-\frac{K^2}{K_0^2}\right)
\ee
We can further simplify Eq.~(\ref{eqn:deltaIt_0}) by noticing that for $t,t_1\gg 1/\Lambda$
$$
\partial_t G_0^R(t-t_1)\simeq -\pi K\delta(t-t_1)
$$
Then  we can write the correction to the time dependent current as
\be\label{eqn:deltaIt}
\delta I(t)=-\tilde{C}\int_0^{t} d\tau\, \frac{\sin\,\left[2K\arctan\Lambda\tau\right]}{\left(1+\Lambda^2\tau^2\right)^{K_{\rm neq}}}
f(t,t-\tau)\,\Phi(t,t-\tau)
\ee
with $\tilde{C}=\pi K C=4e^2KV g_{\rm bs}^2/\pi$. In the next section we will discuss the behavior of this quantity as a function of time and quench amplitude.

\begin{figure}[t]
\begin{center}
\epsfig{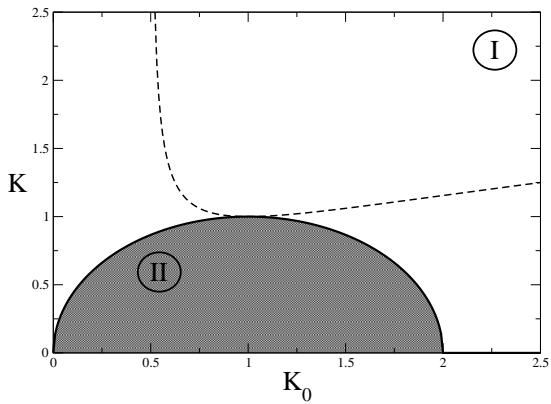}
\caption{Parameter space for the weak-backscattering problem. In the white area (region I) corresponding to $K_{\rm neq}>1$, perturbation theory is
well behaved, while the shaded area (region II) corresponds to the regime where backscattering is a relevant perturbation in the steady state,
and the perturbative expansion breaks down. The dashed line is a reference only to the validity of the strong coupling (weak-tunneling) expansion (see further below).}
\label{fig:fig1}
\end{center}
\end{figure}

\subsection{Discussion}

We start by discussing the equilibrium zero temperature case corresponding to $K_0=K$. In this case we have
$K_{\rm tr}=0$ and $K_{\rm neq}=K$ and the integral in Eq.~(\ref{eqn:deltaIt}) simplifies.
We obtain for the transient current at long times ($\Lambda t\gg1$)
\be
\delta I(t)/V\sim \frac{1}{t^{2(K-1)}}
\ee
i.e., for $K>1$ the weak-backscattering correction to the current vanishes in the long time limit as a power law,
the junction is perfectly conducting and the backscattering is irrelevant, this is the well known result from 
Kane and Fisher~\cite{KaneFisherPRB92}.
As soon as the quench amplitude is non-zero, $K\ne K_0$, we find a number of interesting differences.
We first focus on the long time steady state value of the backscattering current, $\delta I_{\rm ss}=I_{\rm ss}-I_{\rm 0ss}$, this is
\be
\delta I_{\rm ss}/V=
-4\pi K^2 g_{\rm bs}^2 G_0 \int_0^{\infty}d\tau\,\frac{\tau\,\sin\left(2K\arctan\Lambda\tau\right)}{\left(1+\Lambda^2\tau^2\right)^{K_{\rm neq}}}
\ee
$G_0=e^2/2\pi\hbar=e^2/h$ the quantum of conductance, which we set to one from now on, $G_0\equiv 1$.
\begin{figure}[t]
\begin{center}
\epsfig{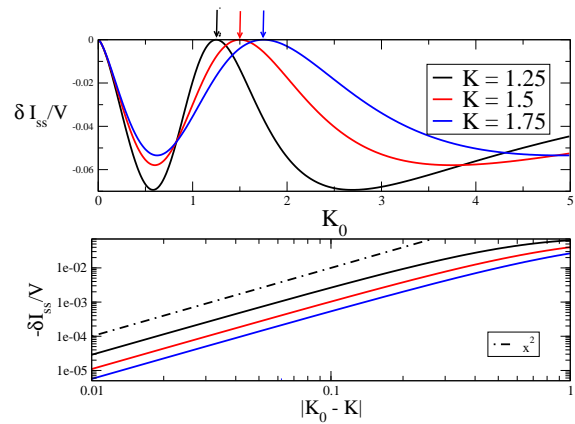}
\caption{Backscattering corrections to steady state current (top panel) as a function of $K_0$ and for different values of $K$. A finite quench amplitude restores transport with deviations from $K_0=K$ (arrows) vanishing quadratically for small quenches, $\delta I_{\rm ss}/V\sim -g_{\rm bs}^2 (K-K_0)^2$ (see bottom panel, log-log scale). Parameters: $\Lambda=1,g_{\rm bs}=0.1$}
\label{fig:fig2}
\end{center}
\end{figure}
We first notice that the integral is well defined as long as $K_{\rm neq}>1$, as expected from perturbative
RG calculations~\cite{SchiroMitraPRL14}, since only in this regime the local backscattering is an irrelevant perturbation for the steady state.
We plot in figure~\ref{fig:fig1} the region of parameters where this condition is satisfied, and our calculations are therefore controlled.
In figure~\ref{fig:fig2} we plot the backscattering correction to the steady state current for different values of $K$ in region I,
as a function of $K_0$. Quite interestingly, the correction to the steady state current is different from zero at any finite quench
amplitude $K\neq K_0$, in other words the systems deviates from the perfect conduction limit.  For a small quench amplitude
one finds that the steady state correction to the conductance is proportional to $(K-K_0)^2$,
\be
\delta I_{\rm ss}/V \sim -g_{\rm bs}^2\left(K-K_0\right)^2
\ee

\begin{figure}[t]
\begin{center}
\epsfig{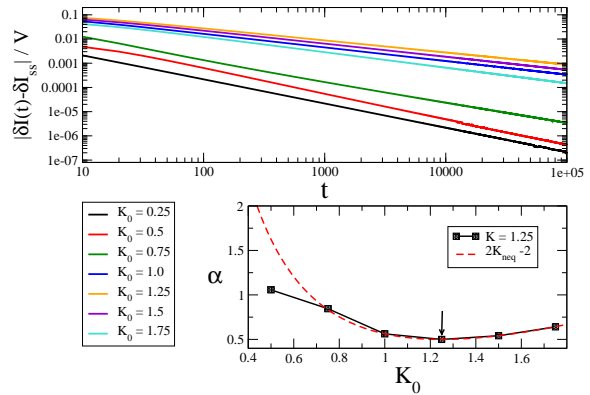}
\caption{Power law decay in time of the weak backscattering correction to the current, for fixed $K=1.25$ and different values of $K_0$. We notice the exponent $\alpha$ depends non-monotonically on $K_0$ and takes a minimum value in the equilibrium case $K=K_0$ (see bottom panel).}
\label{fig:fig3}
\end{center}
\end{figure}

We now discuss the transient behavior and the approach to the steady state. To this extent we evaluate numerically the
integral in Eq.~(\ref{eqn:deltaIt}) and plot the result in figure~\ref{fig:fig3}. We find that the current decays to the
steady state in a power law fashion,
\be
\delta I(t)-\delta I_{\rm ss}\sim \frac{1}{t^{\alpha}}
\ee
with an exponent $\alpha(K,K_0)$ whose dependence on the quench parameters is shown in the bottom panel of figure \ref{fig:fig3},
as a function of $K_0$ at fixed $K=1.25$ in the region I.
We notice that the exponent $\alpha$ behaves non-monotonically with $K_0$ and reaches a minimum value for $K=K_0$, i.e.,
the decay of the current is stronger in the presence of a finite quench amplitude. We can get an analytical understanding of this by looking
at the integral expression for $\delta I(t)$, Eqn.~(\ref{eqn:deltaIt}).
Using the fact that for large $t$ we have $f(t,t-\tau)\sim 1+{\cal O}(1/t)$ while $\Phi(t,t-\tau)\sim
\tau+{\cal O}(1/t)$, we obtain $\alpha=2(K_{\rm neq}-1)$. We notice that this is indeed confirmed by the numerical evaluation of the integral, although for larger values of the bulk quench, deviations start to appear that we attribute to the
finite time resolution of the numerical integration.

We have analyzed only a few values of $K$ while tuning $K_0$, yet we expect this behavior to hold throughout the region I in figure~\ref{fig:fig1}.
We however expect the perturbative expansion to eventually break down as we approach the regime $K_{\rm neq}=1$ and enter region II,
where the backscattering becomes a relevant perturbation, {\sl i.e.}, its strength grows under RG.
In equilibrium, a complementary approach to explore this strong coupling regime amounts to starting from the limit
of two decoupled semi-infinite systems and switching on a local tunneling. In the next section we will suitably generalize
this approach to the quench case and discuss to what extent it can be used to extract useful information about the strong coupling regime.

\section{Transport in the Tunneling Regime}\label{sect:Tunn}

We now consider a different quench protocol which will allow us to access the strong coupling regime where the back-scattering is relevant.
Let us suppose at $t=0$ we have two disconnected and identical semi-infinite TL models describing 1D interacting spinless fermions.
By definition, each lead contains left and right moving fermions interacting in the bulk through intra-branch ($g^0_4$) and inter-branch ($g^0_2$)
scattering processes, as in the previous case. A major difference however arises due to the presence of a sharp edge in each lead,
which imposes open boundary conditions for the fermionic field.
As a result, and differently from the translational invariant case of the previous section,
the two fermionic species are not independent of each other. This suggests an equivalent and more convenient representation of each lead
$i=1,2$ in terms of a single chiral fermionic field $\psi_i(x)$ defined on an infinite system~\cite{FabrizioGogolinPRB95}
({\sl i.e.}, obeying periodic boundary conditions). In terms of this field, the initial Hamiltonian in each lead becomes
\be\label{eqn:H0_i}
H_{0i} =H_{\rm free}+\frac{g_4^0}{2}\,\int dx\,\rho_{i}(x)\rho_{i}(x)
+\frac{g_2^0}{2}\,\int dx\rho_{i}(x)\,\rho_{i}(-x)
\ee
where $H_{\rm free}=v_F\int dx\psi^{\dagger}_{i}i\partial_x\psi_{i}$, while $\rho_{i}(x)=\psi^{\dagger}_{i}(x)\psi_{i}(x)$, and one should notice the non-local
$g_2^0$ interaction that results from the single chiral field representation.

We prepare the system in the ground state $\vert \Psi_0\rangle$ of $H_0=\sum_{i}\,H_{0i}$ and then, for $t>0$, we evolve the system with a different Hamiltonian $H$
\be
H=\sum_{i=1,2}\,H_{i}+\sum_{i=1,2}\,\mu_{i}N_{i}+H_{T}
\ee
where we have (a)  quenched the bulk interactions,
\be\label{eqn:H_i}
H_{i} =H_{\rm free}+\frac{g_4}{2}\,\int dx\,\rho_{i}(x)\rho_{i}(x)
+\frac{g_2}{2}\,\int dx\rho_{i}(x)\,\rho_{i}(-x)
\ee
 (b) switched on a tunneling term coupling the two wires at the edge
\be
H_T =\lambda_0\left(\psi^{\dagger}_1(0) \psi_2(0)+h.c.\right)
\ee
and finally (c) switched on a bias voltage $\mu_{1,2}=\pm eV/2$ that couples to the charge imbalance. We are interested in computing the time dependent current
\be\label{eqn:Itunneling}
I(t)=\langle\Psi(t)\vert\hat{I}\vert\Psi(t)\rangle=\langle\Psi_0\vert\,e^{iHt}\,\hat{I}\,e^{-iHt}\vert\Psi_0\rangle
\ee
where the current operator is
\be
\hat{I}=ie[H,\left(N_1-N_2\right)]=-2ie\lambda_0\left(\psi^{\dagger}_1(0)\psi_2(0)-h.c.\right)
\ee
In order to proceed, it is useful to perform a unitary transformation to eliminate the bias voltage in $H$,
which amounts to going to the rotating frame defined by an operator
$\Omega^{\dagger}(t)=\exp\left(i\sum_{i}\mu_{i}\,N_{i}\,t\right)$. Inserting $\Omega$ into the expression for the current~(\ref{eqn:Itunneling}) we obtain
\be\label{eqn:Itunneling2}
I(t)= \langle\tilde{\Psi}(t)\vert\hat{I}(t)\vert\tilde{\Psi}(t)\rangle
\ee
where the new state
\be
\vert\tilde{\Psi}(t)\rangle=\Omega^{\dagger}(t)\vert\Psi(t)\rangle
\ee
satisfies the Schrodinger equation
\be
i\partial_t\vert\tilde{\Psi}(t)\rangle =
\tilde{H}(t)\vert\tilde{\Psi}(t)\rangle
\ee
with a rotated-frame Hamiltonian
\be
\tilde{H}(t)=\left(i\partial_t\Omega^{\dagger}\right)\Omega+\Omega^{\dagger}H\Omega\equiv
\sum_{i}\,H_{i}+H_{T}(t)
\ee
While the bias voltage has disappeared, the tunneling has become explicitly time-dependent
\be
H_T(t)=\Omega^{\dagger}(t)H_T\Omega(t)=\lambda_0\left(e^{ieVt}\psi^{\dagger}_1(0)\psi_2(0)+h.c.\right)
\ee
where we have used the fact that
\be
\Omega^{\dagger}(t)\psi_{i}(x)\Omega(t)=e^{-i\mu_{i}t}\psi_i(x)
\ee
The current operator in~(\ref{eqn:Itunneling}) also acquires an explicit time dependence
\be
\hat{I}(t)= \Omega^{\dagger}(t)\hat{I}\Omega(t)=-2ie\lambda_0\left(
e^{ieVt}\psi^{\dagger}_1(0)\psi_2(0)-h.c.\right)
\ee
The time dependent state $\vert\tilde{\Psi}(t)\rangle$ can be written in the interaction picture $O_I = e^{iH_i t}Oe^{-i H_i t}$
with respect to the unperturbed Hamiltonian of the uncoupled wires as
\be
\vert\tilde{\Psi}(t)\rangle=e^{-i\sum_iH_{i} t}T\,e^{-i\int_0^t\,dt'\,H_{T,I}(t')}\vert\Psi_0\rangle
\ee
We can now evaluate the time dependent current to the lowest order contribution in the tunneling, which gives
\be
 I(t)=-i\int_0^t\,dt_1\,\langle\Psi_0\vert\left[\hat{I}_I(t),H_{T,I}(t_1)\right]\vert\Psi_0\rangle
\ee
where all the operators are explicitly time dependent as in previous equations and evolved with the
Hamiltonian of the uncoupled wires, after the global quench.

%\subsection{Open Boundary Bosonization Approach}\label{sect:cond_tun_boson}

It is now useful to use bosonization, in its open-boundary formulation~\cite{FabrizioGogolinPRB95} to proceed with the evaluation of the tunneling current. We first write the fermionic operator in each lead in terms of a single chiral bosonic mode
\be
\psi_{i}(x)=\frac{1}{\sqrt{\alpha}}\,e^{i\varphi_{i}(x)}
\ee
as well as the electron density in the lead $i$ as
\be
\rho_i(x) = \partial_x\varphi_i(x)/2\pi
\ee
The Hamiltonian before and after the global quench of the interactions (Eqns~\ref{eqn:H0_i}  and \ref{eqn:H_i}) become now quadratic in this bosonic field and can be easily diagonalized with a Bogoliubov rotation (see appendix~\ref{sect:quench_open_bound}). For example we have for $H_{0i}$
\bea
H_{0i} =\frac{u_0}{4}\left(\frac{1}{K_0}+K_0\right)\int\frac{dx}{2\pi}(\partial_x\varphi_i(x))^2+\nonumber\\
-\frac{u_0}{4}\left(\frac{1}{K_0}-K_0\right)\,\int\frac{dx}{2\pi}\partial_x\varphi_i(x)\partial_x\varphi_i(-x)
\eea
The resulting harmonic theory is again described in terms of two parameters, the sound velocity $u_0,u$ and the Luttinger parameter
$K_0,K$, which by construction are identical in the two leads, and whose expressions in terms of the coupling constant
$g^0_2,g^0_4$ and $g_2,g_4$ are the same as in the translational invariant case (see equations~(\ref{eqn:u0K0})). We rewrite them here for convenience
\bea
u_0 =v_F\,\sqrt{\left(1+g^0_4/2\pi v_F\right)^2-(g^0_2/2\pi v_F)^2}\\
K_0 = \sqrt{\frac{1+g^0_4/2\pi v_F-g^0_2/2\pi v_F}{1+g^0_4/2\pi v_F+g^0_2/2\pi v_F}}
\eea
with similar relations holding for $u,K$ as functions of $g_2,g_4$.  In addition  we need to write the tunneling and current operator in the bosonic language. To this extent it is useful to introduce the combinations
\be
 \varphi^{\pm}(x)=\frac{\varphi_1(x)\pm\varphi_2(x)}{\sqrt{2}}
\ee
and to notice that only the $\varphi^-$ combination enters both the tunneling and the current.
Indeed we have disregarded the Klein factors which can be shown to be unimportant. Thus,
\bea\label{eqn:tunn_bosons}
H_T(t)= 2\lambda \cos\left(\sqrt{2}\varphi^-(t)-eVt\right)\\
\label{eqn:curr_bosons}
\hat{I}(t)=-4e\lambda\sin \left(\sqrt{2}\varphi^-(t)-eVt\right)
\eea
where we have defined $\varphi^-(t)\equiv\varphi^-(0,t)$ and $\lambda=\lambda_0/\alpha$.

\subsection{Tunneling Correction To Current}\label{sect:tunn_correction}

In order to proceed further, we plug Eqns.~(\ref{eqn:tunn_bosons},\ref{eqn:curr_bosons})  in the expression for the tunneling current  to obtain, after some simple algebra,
\be\label{eqn:Itunn}
  I(t)=-8\,e\lambda^2\,\int_0^t\,d\tau\,\sin\left(eV\tau\right)\,F^R(t,t-\tau)
\ee
where $F^R(t,t')$ is the retarded component of the correlator
\bea
F_{ab}(t,t')&=&-i\langle \cos\left(\sqrt{2}\varphi^-_{a}(t)\right)\cos\left(\sqrt{2}\varphi^-_{b}(t')\right)\rangle
=\nonumber\\
&=&-\frac{i}{2}\langle\,e^{i\sqrt{2}\varphi^-_{a}(t)}\,e^{-i\sqrt{2}\varphi^-_{b}(t')}\rangle\,
\eea
It is convenient to express this correlator in terms of the fields $\varphi_{i}(t)$ in the two semi-infinite leads. We have
\bea
F_{ab}(t,t')=
-\frac{i}{2}\langle\,e^{i\varphi_{1a}(t)}\,e^{-i\varphi_{1b}(t')}\rangle\,
\langle\,e^{i\varphi_{2b}(t')}\,e^{-i\varphi_{1a}(t)}\rangle\nonumber
\eea
where we have used the fact that in the absence of tunneling the two leads are decoupled. To compute this correlator we need therefore
the local Green's function of the field $\varphi_{i}$ after a quench of the bulk interaction parameter in a semi-infinite chain,
which we give in appendix~\ref{sect:quench_open_bound}. The result for the retarded component, $F^R=-\sum_{ab}b\,F_{ab}/2$ is found to be
\be
F^R(t_1>t_2)=
-\frac{\sin\left(\frac{2}{K}\arctan\Lambda(t_1-t_2)\right)}
{\left(1+\Lambda^2\left(t_1-t_2\right)^2\right)^{K^{\rm dual}_{\rm neq}}}
f(t_1,t_2)
\ee

\begin{figure}[t]
\begin{center}
\epsfig{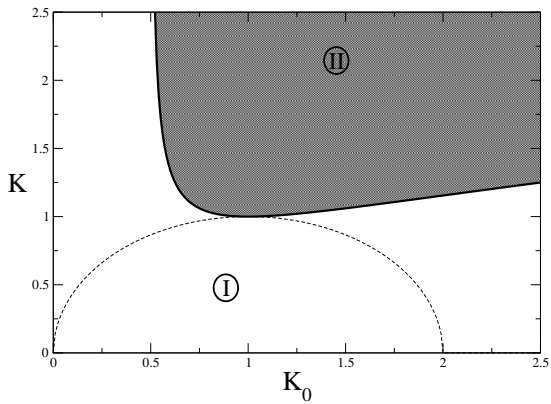}
\caption{Parameter space for the tunneling problem. In the white area (region I) corresponding to $K^{\rm dual}_{\rm neq}>1$, perturbation theory is well behaved, while the shaded area (region II) corresponds to the regime where the tunneling is a relevant perturbation in the steady state and the perturbative expansion breaks down. The dashed line is a reference only to the validity of the weak backscattering expansion previously discussed.}
\label{fig:fig4}
\end{center}
\end{figure}
where we have introduced
\be
f(t_1,t_2)=\left[
\frac{\left(1+4\Lambda^2 t_1^2\right)\left(1+4\Lambda^2 t_2^2\right)}
{\left(1+\Lambda^2(t_1+t_2)^2\right)^2}\right]^{K^{\rm dual}_{\rm tr}/2}
\ee
with the exponents
\be\label{eqn:KneqKtrdual}
K_{\rm neq}^{\rm dual}=\frac{1}{2K_0}\left(1+\frac{K_0^2}{K^2}\right)\qquad
K_{\rm tr}^{\rm dual}=\frac{1}{2K_0}\left(1-\frac{K_0^2}{K^2}\right)
\ee
We stress that Eq.~(\ref{eqn:Itunn}) is perturbative in the tunneling but contains the bias voltage $eV$ to all orders. 
In fact in the absence of the bulk quench, $K=K_0$, we can recover from this the well known result~\cite{KaneFisherPRB92} 
for the non-linear I-V characteristic of the wire, $I(V)\sim V^{2/K-1}$. 
In the following we will instead consider mostly the low bias regime, for a finite quench amplitude and to this extent we 
will evaluate the current in the low bias regime and equivalently the non-linear differential conductance at zero bias.
Note that the latter $I/V|_{V=0}=0$ in the absence of quench. 

After some simple manipulations, we can write the time dependent current at small bias voltage as
\be\label{eqn:Itun_final}
I(t)= 8e^2V\lambda^2\,\int_0^t\,d\tau\,
\frac{\tau\sin\left(\frac{2}{K}\arctan\Lambda\tau\right)}
{\left(1+\Lambda^2\tau^2\right)^{K^{\rm dual}_{\rm neq}}}
f(t,t-\tau)
\ee

We notice the result we have obtained for the current in the tunneling regime is \emph{dual} to the one in the weak-backscattering case
in the sense that the two currents are related by a transformation of the Luttinger parameters, before and after the quench, into
\be
K,K_0\longrightarrow \frac{1}{K},\frac{1}{K_0}
\ee
This is the natural generalization of the equilibrium duality~\cite{Giamarchi_2003}. Yet, as one can immediately see from the
expressions~(\ref{eqn:KneqKtr}) for $K_{\rm neq},K_{\rm tr}$ and (\ref{eqn:KneqKtrdual}) for $K_{\rm neq}^{\rm dual}, K_{\rm tr}^{\rm dual}$,
the exponents controlling the decay of the correlation in the tunneling and weak-backscattering regime are not related by this simple duality,
{\sl i.e.},
\bea\label{eqn:noduality1}
K_{\rm neq}^{\rm dual}\neq 1/K_{\rm neq}\\
\label{eqn:noduality2}
K_{\rm tr}^{\rm dual}\neq 1/K_{\rm tr}
\eea
unless of course $K=K_0$. This will have important consequences as we are going to discuss in the next section.
A simple way to understand the origin of this result is to notice that while for the Hamiltonian after the quench the duality still holds, 
the initial condition of the problem (the ground state of the Luttinger model with interaction parameter $K_0$) is not dual when written in 
the basis of the natural eigenmodes of the system after the quench (Luttinger model with interaction parameter $K$). 
The fact that the Luttinger model is integrable and therefore never fully loses memory of this initial condition implies 
the breakdown of duality in the stationary state.

\subsection{Discussion}

As in the weak-backscattering case, we start discussing the equilibrium zero temperature result corresponding to $K_0=K$. In this case we
have $K_{\rm tr}=0$ and $K_{\rm neq}=K$, and the integral in
Eq.~(\ref{eqn:Itun_final}) simplifies. For the transient current at long times ($\Lambda t\gg1$) we obtain,
\be
I(t)/V\sim \frac{1}{t^{1/K-1}}
\ee
i.e. for $K<1$ the correction to the current vanishes in the long time limit as a power law, the junction is perfectly insulating and the tunneling is irrelevant~\cite{KaneFisherPRB92}.
\begin{figure}[t]
\begin{center}
\epsfig{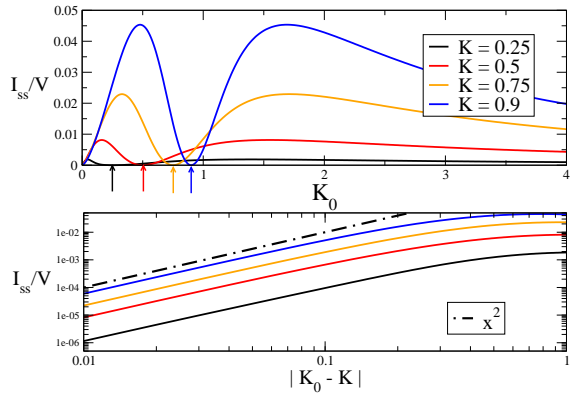}
\caption{Steady state tunneling current as a function of $K_0$ and for different values of $K$ (top panel). We see that a non-zero quench amplitude results in a finite tunneling current, vanishing quadratically as $K_0\rightarrow K$ (arrows),
$ I_{\rm ss}/V\sim \lambda^2 (K-K_0)^2$ (see bottom panel, log-log scale). Parameters: $\Lambda=1,\lambda=0.1$ }
\label{fig:fig5}
\end{center}
\end{figure}
We now consider the finite quench case, $K\ne K_0$, and first focus on the long time steady state value of the tunneling current which reads
\be
 I_{\rm ss}/V = 8 e^2\lambda^2\,\int_0^{\infty}\,d\tau\,\frac{\tau \sin\left(\frac{2}{K}\arctan\Lambda\tau\right)}{\left(1+\Lambda^2\tau^2\right)^{K_{\rm neq}^{\rm dual}}}\,
\ee
We notice that the integral is well defined as long as $K^{\rm dual}_{\rm neq}>1$, a condition which is satisfied in the region of parameters plotted in figure~\ref{fig:fig4} (region I). Interestingly, this region is not simply the dual of the region I for the weak-backscattering case, (light shaded line inside region I in figure~ \ref{fig:fig4}) as it would be in the equilibrium case. This is a consequence of Eqns~(\ref{eqn:noduality1},\ref{eqn:noduality2}). Therefore we conclude that the quenched impurity model does not display a full duality, contrarily to other nonequilibrium realizations such as the driven noisy one~\cite{DallaTorreetalPRB2012}.  We will come back to this point in the next section.
%One interesting consequence of this result is that there are regions of parameter space where neither the backscattering, nor the tunneling are relevant perturbations.
Let us now discuss the behavior of the steady state current. Quite interestingly, the steady state current is different from zero at any finite quench amplitude $K\neq K_0$, in other words the system deviates from the perfect insulating limit. In figure~\ref{fig:fig5} we plot the steady state value of the tunneling current at fixed $K$ and as a function of $K_0$. As in the weak-backscattering limit we find that for small quench amplitudes the steady state current is
\be
I_{\rm ss}/V\sim \lambda^2(K-K_0)^2
\ee

\begin{figure}[t]
\begin{center}
\epsfig{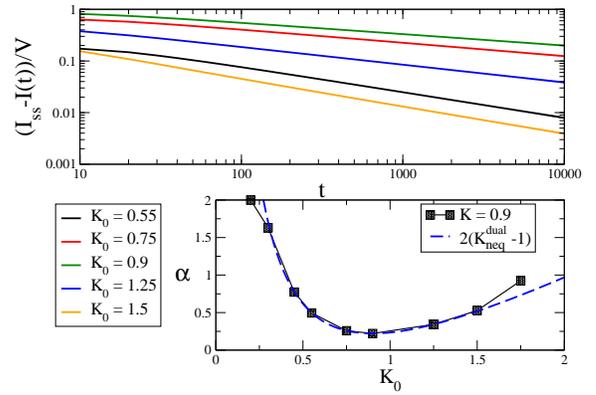}
\caption{
Power law decay in time of the weak tunneling correction to the current, for fixed $K=0.9$ and different values of $K_0$. We notice the exponent $\alpha$ depends non-monotonically on $K_0$ and takes a minimum value in the equilibrium case $K=K_0$ (see bottom panel).
%Steady state current versus Luttinger parameter $K_0$, for fixed quench amplitude $K-K_0$, both in the tunneling regime (left panel) and in the weak-backscattering one. In equilibrium a sharp transition between an insulating link (for $K<1$) and a conducting junction (for $K>1$) exists, which turns into a crossover in the presence of a non-vanishing quench amplitude.
}
\label{fig:fig6}
\end{center}
\end{figure}

We now discuss the transient behavior and the approach to the steady state. We evaluate numerically the integral in eq.~(\ref{eqn:Itun_final}) 
and plot the result in figure~\ref{fig:fig6}. We find that the current decays to the steady state as a power law
\be
 I(t)- I_{\rm ss}\sim  \frac{1}{t^{\alpha}}
\ee
with an exponent $\alpha(K,K_0)$ whose dependence on the quench parameters is shown in figure~\ref{fig:fig6}. As in the weak-backscattering case previously discussed, we find that the decay of the current is in general faster for a finite quench amplitude, in accordance with the analysis of Ref.~\onlinecite{PerfettoStefanucciCiniPRL10}. In particular the exponent $\alpha$ reaches its minimum for $K=K_0$ and behaves as $\alpha\sim 2(K_{\rm neq}^{\rm dual}-1)$.

\section{Discussion}

Let us summarize the results of previous sections and discuss their physical consequences. We started from the limit of a uniform system and found that for $K_{\rm neq}>1$, corresponding to region I of fig.~\ref{fig:fig1}, perturbation theory in the backscattering is well behaved. While this means that in equilibrium, $K\rightarrow K_0$, the conductance approaches the unitary limit, a bulk quench $K\neq K_0$ gives rise to a finite deviation from this perfect conducting regime. This is consistent with the RG analysis we developed in Ref.~\onlinecite{SchiroMitraPRL14}. There we showed that although for $K_{\rm neq}>1$ the backscattering is an irrelevant perturbation for the steady state, it nevertheless gives a sizeable effect, generating an effective temperature for the local degree of freedom, which explains, at least qualitatively, the deviation from the unitary limit. When $K_{\rm neq}<1$, the backscattering becomes relevant, and perturbative approaches in the back-scattering breaks down, and one enters the strong coupling regime.

A possible approach to grasp the behavior of the system in this limit of strong backscattering is to start from the strong coupling fixed point, assuming the effective temperature is not enough to cut the growth of the backscattering under RG flow, and perturb this system of two decoupled wires with the sudden switching on of a local tunneling. We have considered this regime in section~\ref{sect:Tunn} and found that for $K^{\rm dual}_{\rm neq}>1$, corresponding to region I in fig.~\ref{fig:fig4}, perturbation theory in the tunneling is well behaved. While this could naively suggest the strong coupling fixed point remains stable as in equilibrium, we have found that a finite tunneling current appears in the steady state for any finite quench amplitude $K\neq K_0$. This result is again reminiscent of a thermal behavior and we may speculate that an effective temperature would indeed be generated under RG by the local tunneling, in analogy with the backscattering case.

Taken together, these results suggest that a global bulk quench has a dramatic effect on the problem, smearing out the sharp distinction between the conducting and the insulating phase that exists in equilibrium at zero temperature.
\begin{figure}[t]
\begin{center}
\epsfig{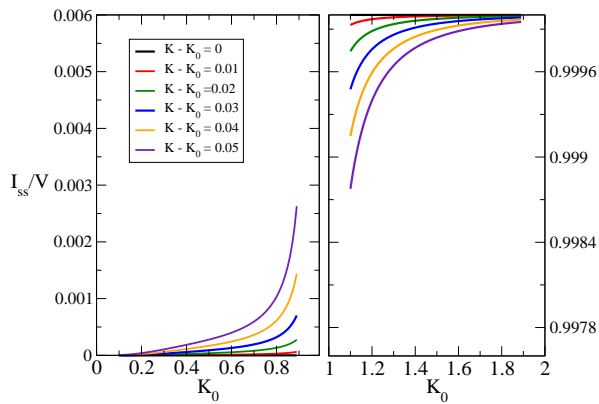}
\caption{Steady state current versus Luttinger parameter $K_0$ at fixed values of the quench amplitude $K-K_0$, in the tunneling regime (left panel, for $\lambda=0.1$) and in the weak backscattering limit (right panel, for $g_{\rm bs}=0.1$). We see that the sharp transition between insulator and conductor driven by $K_0$ gets smeared out by the global quench.}
\label{fig:fig7}
\end{center}	
\end{figure}
To further appreciate the role played by the quench amplitude in the problem, it is useful to look again at the steady state current in both regimes. In figure~\ref{fig:fig7} we plot the weak-tunneling and weak-backscattering correction to the current as a function of $K$ and at fixed $K-K_0$. At zero quench amplitude this would give the usual sharp transition from an ideal insulator for $K<1$ to a perfect conductor for $K>1$. However as we clearly see, any small finite quench amplitude turns this into a smooth crossover. Finally, it is worth noticing that, as we mentioned earlier, the two regions $K_{\rm neq}>1$ and $K_{\rm neq}^{\rm dual}>1$ are not dual to each other. This means that there exist regions of the parameter space $(K,K_0)$ where neither the backscattering nor the tunneling are relevant, i.e., perturbation theory is well behaved. We interpret this result as a further signature of the effective temperature behavior of the problem.

These results for transport offer therefore another example of the quench-induced decoherence mechanism that we had previously identified in the orthogonality catastrophe problem~\cite{SchiroMitraPRL14}.
As in that case, the relevant energy scale, in the limit of small quenches, is proportional to the quench amplitude squared
\be
\gamma_{\star}\sim g^2\left(K-K_0\right)^2
\ee
where $g$ in the equation above indicates any source of local non-linearity, either due to the backscattering potential in the weak impurity limit, or 
due to the tunneling term in the limit of two decoupled chains. 
While the qualitative behavior is suggestive of an effective thermal behavior, it is important to stress that such an analogy does not fully carry over. In particular, the characteristic power law structure exhibited by the finite temperature current in the equilibrium problem~\cite{KaneFisherPRL92,KaneFisherPRB92,Giamarchi_2003}, on both sides of the transition, appears to be washed out by nonequilibrium effects. Indeed, as we have discussed earlier, the scaling of the current in the steady state does not show signature of non-trivial power laws, even in the small quench amplitude regime where transport is essentially set by the quench-induced decoherence scale $\gamma_{\star}$, with very weak additional dependence from the Luttinger parameters.  This suggests that the analogy between this scale and an effective temperature, meant here as an infrared cutoff, cannot be pushed too far.
 
 Finally it is worth comparing our results for the steady state impurity physics with the results and the predictions
of Ref.~\onlinecite{KennesMedenPRB13}. Here the authors computed the zero frequency charge susceptibility in a uniform LL after an interaction quench starting from a free system. For repulsive interactions they found it to diverge
at the backscattering wave vector and concluded that, as in equilibrium, even a weak single impurity
strongly disturbs the homogeneous LL or, in an RG language, the perfect chain fixed point is unstable.
Then, they computed the local density of states close to an open boundary and found that it saturates
to a non-zero value, with power-law corrections at low frequency.
From the scaling of these corrections they conclude that the steady state analog of the open-chain fixed
point is stable, with a power-law finite temperature conductance, which eventually crosses over to a non-zero value at low
enough temperature. Based on their results the authors argue that the fixed point structure of a single impurity
in a nonequilibrium steady state LL is similar to the one in equilibrium~\cite{KennesMedenPRB13}.

As we have seen, our transport results combined with our analysis of the transient orthogonality catastrophe
problem~\cite{SchiroMitraPRL14}, suggest a rather different picture and highlight
the importance of inelastic effects for the steady state impurity physics after the quench.
In particular in the back-scattering case we found that, due to the global quench, a local non-linearity generates
an effective decoherence in the problem, even in the regime where it would nominally be irrelevant at the one
loop level. This \emph{quench-induced decoherence} energy scale reduces the value of the zero temperature conductance
away from the perfect uniform limit and, more importantly, competes with the growth of the backscattering as
one enters the strong coupling phase, corresponding to $K < 1$ for the case $K_0 = 1$ considered in Ref.~\onlinecite{KennesMedenPRB13}.
The result of this competition is clearly a subtle issue to establish firmly. Our results for the Loschmidt echo~\cite{SchiroMitraPRL14} 
suggested that well within the putative strong coupling phase this new energy scale acts as an effective infrared cut off, 
resulting in an exponential decay of the echo.
This result would suggest that, quite differently from the equilibrium case, a weak single impurity does not disturb
substantially the uniform system, i.e., one should not trust the perturbative break down for $K_{\rm neq} < 1$.

To further confirm this intuition, we have considered the opposite limit of an open chain, perturbed by
an interaction quench and by a sudden switching on of a local tunneling. For any finite bulk quench amplitude we
find a non-vanishing zero temperature conductance essentially set by the quench-induced decoherence scale.
This result suggests that, differently from the equilibrium case, one never reaches the open-chain fixed point when the 
bulk has been quenched because the weak tunneling is always effective
in generating a non-zero conductance. We may interpret this in an RG picture for the
tunneling problem, dual to what we discussed in Ref.~\onlinecite{SchiroMitraPRL14} for the backscattering case, where the local non-linearity generates an effective decoherence that cuts off the RG flow.
While a non-zero tunneling conductance was also
found in Ref.~\onlinecite{KennesMedenPRB13}, the physics behind this effect was not discussed. 
We now fully clarify its origin as a quench-induced decoherence phenomenon.

\section{Conclusions}

In this paper we have studied the problem of transport through a localized impurity in a Luttinger model far from equilibrium due to a
bulk interaction quench. This model in its ground state has a familiar and well studied zero temperature transition between a conducting and an insulating phase, depending on the value of the bulk interaction parameter $K$, and has an associated
finite-temperature and finite-voltage correction which shows universal power law behavior.

Here we have presented perturbative calculations in the strength of the impurity potential
starting both from the limit of a uniform liquid and from the one of two decoupled semi-infinite systems. In the former case a standard bosonization approach gives rise to a local backscattering term (described by a boundary sine-Gordon problem in a time-dependent nonequilibrium bath) that we treat to leading order in perturbation theory. In the latter, we use open boundary bosonization to treat the problem and obtain a dual formulation again in terms of a boundary sine-Gordon
model where the non-linearity comes from a local tunneling between the isolated wires. Our results quite generically reveal that the nonequilibrium excitation of bulk modes induced by the global quench has important and peculiar effects on the conducting-insulating transition, which gets smeared out into a crossover for any small quench amplitude $K-K_0\neq 0$. In addition, the dynamics of the current displays a faster decay towards the steady state as compared to the equilibrium zero temperature case.

All together this suggests that the global quench effectively induces a decoherence mechanism for the local degrees of freedom and we have identified
an energy scale  $\gamma_{\star}$ associated with this. While this behavior is qualitatively similar to a finite effective temperature, the decoherence energy scales enters the problem differently than a temperature. In particular the steady state current in both limits is set by the decoherence scale but does not show power laws in this energy scale.

An interesting question concerns the generalities of these results beyond the quenched Luttinger model. We may speculate that quench-induced decoherence is a generic phenomenon that occurs in other interacting quantum impurity problems coupled to environments which are non-thermal at long times but rather flow to a Generalized Gibbs ensemble steady state. We leave the investigation of this intriguing question to future work.

\section{Acknowledgment}

We thank Ehud Altman, Volker Meden, Achim Rosch and Hubert Saleur for helpful discussions. AM was supported by NSF-DMR 1303177.

\appendix

\section{Green's Functions -- Definitions and Identities}\label{sect:GF}

We define the contour ordered Green's function for a real bosonic field $\phi(t)$ in the $\alpha,\beta=\pm$ Keldysh basis
\be
G^{\alpha\beta}(1,2)=-i\langle\,\phi_{\alpha}(1)\phi_{\beta}(2)\rangle
\ee
where we have, following the convention of Ref.~\onlinecite{TavoraAditiPRB14}
\bea
G^{--}(1,2)\equiv G^T(1,2)= -i\langle\,T\phi(1)\phi(2)\rangle\\
G^{-+}(1,2)\equiv G^<(1,2)= -i\langle\,\phi(2)\phi(1)\rangle\\
G^{+-}(1,2)\equiv G^>(1,2)= -i\langle\,\phi(1)\phi(2)\rangle\\
G^{++}(1,2)\equiv G^{\tilde{T}}(1,2)= -i\langle\,\tilde{T}\phi(1)\phi(2)\rangle
\eea
We then define the retarded, advanced and Keldysh components as
\bea
G^R(1,2)=-i\theta(1-2)\langle [\phi(1),\phi(2)]\rangle\\
G^A(1,2)=i\theta(2-1)\langle [\phi(1),\phi(2)]\rangle\\
G^K(1,2)=-i\langle \left\{\phi(1),\phi(2)\right\}\rangle
\eea
and find the following useful relation
\be
2G_{\alpha\beta}(1,2)= G^K(1,2)-\alpha G^A(1,2)-\beta G^R(1,2)
\ee
from which the following relations follow
\bea \label{eqn:id1}
G^R(1,2)=-\frac{1}{2}\sum_{\alpha\beta}\,\beta G_{\alpha\beta}(1,2)\\
 \label{eqn:id2}
G^K(1,2)-\beta G^R(1,2)=\sum_{\alpha}\, G_{\alpha\beta}(1,2)\\
 \label{eqn:id3}
G^K(1,2)=\frac{1}{2}\sum_{\alpha\beta}\, G_{\alpha\beta}(1,2)
\eea

\section{Forward Scattering Contribution to the Current}\label{sect:fwd_scatt}

In this appendix we show that the forward scattering does not contribute to the time dependent current. To see this we follow standard steps~\cite{VonDelftSchoeller_review98,GogolinNerseyanTsvelik_2004} and introduce first the even and odd combinations of the LL fields $\theta,\phi$, defined as
\bea
\phi_{\rm e/o}=\frac{\phi(x)\pm\phi(-x)}{\sqrt{2}}\\
\theta_{\rm e/o}= \frac{\theta(x)\pm\theta(-x)}{\sqrt{2}}
\eea
Then, it is convenient to introduce new chiral bosonic fields $
\Phi_{\rm s/a}(x)=\sqrt{K}\theta_{\rm o/e}(x)+\frac{1}{\sqrt{K}}\phi_{\rm e/o}(x)$
satisfying $\left[\Phi_{\rm s}(x),\Phi_{\rm s}(y)\right]=[\Phi_{\rm a}(x),\Phi_{\rm a}(y)]=-i\pi\,\mbox{sign}(x-y)\,$ and $
\left[\Phi_{\rm s}(x),\Phi_{\rm a}(y)\right]=0$. In terms of these fields the bulk LL Hamiltonian decouples in the two channels
\be
\m{H}=\frac{u}{4\pi} \int\,dx\,\sum_{\nu=s,a}\,\left(\partial_x\Phi_{\nu}\right)^2\equiv \sum_{\nu=s,a}\,\bar{\m{H}}_{\nu}
\ee
It is particularly convenient to write the local scattering in terms of these fields, due to their well defined properties under inversion. We have
\be
V_{\rm loc}=g_{\rm fs}\,\sqrt{\frac{K}{2}}\,\partial_x\,\Phi_{\rm a}(x)\vert_{x=0}+
g_{\rm bs}\,\cos\,\sqrt{2K}\Phi_{\rm s}(x=0)\nonumber
\ee
i.e. the forward scattering couples to the antisymmetric mode while the backward term only to the symmetric one.
Now we recall Eq.~(\ref{eqn:current_wbs}) of the main text, which shows that the time-dependent current $I(t)$ is expressed only in terms of the Green's function of the local field $\phi(0)$, which turns out to be related only to the symmetric combination
$\phi(0)=\sqrt{K/2}\Phi_s(0)$, from which we conclude that the current does not depend on the forward scattering potential.

\section{Quenches in the Open-Boundary Tomonaga Luttinger Model}\label{sect:quench_open_bound}

In this appendix we briefly discuss the quench problem in a TL model with open boundary conditions, and compute in particular the local
bosonic correlator relevant for obtaining the tunneling current. We first recall how to diagonalize the initial Hamiltonian $H_0$,
as the same strategy will be used for the Hamiltonian after the quench, and this will highlight the major differences with the bulk case.
We follow the treatment in Ref.~\onlinecite{FabrizioGogolinPRB95} to which we refer the reader to for further details.

Let us start with the problem of spinless fermions $\psi(x)$ defined on a line $[0,L]$ with open boundary conditions (OBC) with the Hamiltonian
\begin{eqnarray}
&&H_0=H_{\rm free}+\frac{g_4^0}{2}\int_0^L dx\left(\rho_L^2(x)+\rho_R^2(x)\right)\nonumber\\
&&+g_{2}^0\int_0^L dx\rho_L(x)\rho_R(x)
\end{eqnarray}
where $H_{\rm free}=v_F\sum_{\alpha=L,R}s_{\alpha}\int_0^L dx\psi^{\dagger}_{\alpha}i\partial_x\psi_{\alpha}$ with $s_{L}=-s_R=1$. The crucial observation is 
that imposing open boundary conditions introduces a constraint between right and left moving fermions which are no longer independent, 
but rather satisfy~\cite{FabrizioGogolinPRB95}
\be\label{eqn:constraint}
\psi_L(-x)=-\psi_R(x)
\ee
and is a direct consequence of the existence of a single Fermi point. We can therefore fold the line into $(-L,L)$ and use Eq.~(\ref{eqn:constraint}) to express the Hamiltonian only in terms of right movers
\be\label{eqn:H0_i_app}
H_{0} =H_{\rm free}+\frac{g_4^0}{2}\int dx\,\rho_{R}(x)\rho_{R}(x)
+\frac{g_2^0}{2}\int dx\rho_{R}(x)\rho_{R}(-x)
\ee
where $H_{\rm free}=v_F\int dx\psi^{\dagger}_{R}i\partial_x\psi_{R}$ and one has to notice the non-local $g_2^0$ interaction that results from the single chiral field representation. The right moving fermions now satisfy periodic boundary conditions and can be bosonized as usual in terms of a single bosonic field
\be
\psi(x)=\frac{1}{\sqrt{\alpha}}\,e^{i\varphi(x)}
\ee
in terms of which the Hamiltonian reads
\bea
H_{0} =\frac{v_F}{2}\left(1+\frac{g_4^0}{2\pi v_F}\right)\int\frac{dx}{2\pi}(\partial_x\varphi_i(x))^2+\nonumber\\
-\frac{g_2^0}{4\pi}\,\int\frac{dx}{2\pi}\partial_x\varphi_i(x)\partial_x\varphi_i(-x)
\eea
To diagonalize this problem it is useful to decompose the field into normal modes
\be
\varphi(x)=\sum_{q>0}\,\sqrt{\frac{\pi}{qL}}\,e^{-aq/2}\,\left(b_q\,e^{iqx}+b^{\dagger}_q\,e^{-iqx}\right)
\ee
in terms of which the Hamiltonian becomes
\be
H_0=\sum_{q>0}\,A_0q\,b^{\dagger}_q\,b_q-B_0q\,\left(b^{\dagger}_qb^{\dagger}_q+b_q b_q\right)
\ee
where
\bea
A_0 = v_F\left(1+g^0_{4}/2\pi v_F\right)\\
B_0 = \frac{g^0_2}{4\pi}
\eea
We can diagonalize the Hamiltonian via a Bogoliubov rotation parameterized by the angle $\alpha_0$.
\bea
b_q = \cosh\alpha_0\,\eta_{q}-\sinh\alpha_0\,\eta^{\dagger}_{q}\\
b^{\dagger}_q = -\sinh\alpha_0\,\eta_{q}+\cosh\alpha_0\,\eta^{\dagger}_{q}
\eea
The Hamiltonian when written in terms of the fields $\eta_{q}$ is,
\be
H_0= \sum_q\,v_0q\,\eta^{\dagger}_{q}\,\eta_{q}
\ee
for an angle $\alpha_0$ given by
\be\label{eqn:bogolubov_angle}
\tanh 2\alpha_0=-\frac{2B_0}{A_0} =-\frac{g^0_2/2\pi v_F}{1+g^0_4/2\pi v_F}
\ee
The sound velocity reads
\bea
v_0=A_0\,\cosh\,2\alpha_0+2B_0\,\sinh\alpha_0 =\nonumber\\
=v_F\sqrt{\left(1+g^0_4/2\pi v_F\right)^2-\left(g^0_2/2\pi v_F\right)^2}
\eea
while the angle $\alpha_0$ is defined as
\be\label{eqn:K_angle}
\exp\,2\alpha_0=\sqrt{\frac{1+g^0_4/2\pi v_F-g^0_2/2\pi v_F}{1+g^0_4/2\pi v_F+g^0_2/2\pi v_F}}\equiv K_0
\ee
%It is important to stress that while $K_0$ and $v_0$ depends on the microscopic interactions $g^0_2,g^0_4$ as in the bulk problem, there is an important difference in sign in Eq.(\ref{eqn:K_angle}), defining the relation between $K_0$ and $\alpha_0$.

We can now proceed to discuss the  problem of an interaction quench of the bulk interactions $g^0_2,g^0_4\rightarrow g_2,g_4$. This amounts to suddenly change $A_0,B_0\rightarrow A,B$ in the bosonic Hamiltonian. We can repeat the previous steps and introduce operators $\gamma^{\dagger}_q,\gamma_q$
\bea
b_q = \cosh\alpha_0\,\gamma_{q}-\sinh\alpha_0\,\gamma^{\dagger}_{q}\\
b^{\dagger}_q = -\sinh\alpha_0\,\gamma_{q}+\cosh\alpha_0\,\gamma^{\dagger}_{q}
\eea
giving a quadratic Hamiltonian when written in terms of the fields $\gamma_{q}$
\be
H= \sum_q\,vq\,\gamma^{\dagger}_{q}\,\gamma_{q}
\ee
for an angle $\alpha$ given by
\be\label{eqn:bogolubov_angle2}
\tanh 2\alpha=-\frac{2B}{A} =-\frac{g_2/2\pi v_F}{1+g_4/2\pi v_F}
\ee
such that
\be
K=\exp2\alpha
\ee
Using the above transformations, we can easily express the time evolved bosonic operators in terms of those diagonalizing the initial Hamiltonian
\bea
b_q(t) = e^{iHt}\,b_q\,e^{-iHt}= f_q(t)\,\eta_{q}-g_q(t)\,\eta^{\dagger}_{q}\\
b^{\dagger}_q(t) = e^{iHt}\,b^{\dagger}_q\,e^{-iHt} = -g^*_q(t)\,\eta_{q}+f^*_q(t)\,\eta^{\dagger}_{q}
\eea
where (assuming as usual a Galilean invariance preserving quench such that $v_0K_0=vK$)
\bea
f_q(t)=\cosh\alpha_0\,\cos \eps_q t-i\cosh\left(2\alpha-\alpha_0\right)\,\sin\eps_q\,t\nonumber\\
g_q(t)=\sinh\alpha_0\,\cos \eps_q t+i\sinh\left(2\alpha-\alpha_0\right)\,\sin\eps_q\,t\nonumber
\eea
In particular we can write down the time-evolved local bosonic field $\varphi(x=0,t)$ as
\be\label{eqn:td_field}
\varphi(t)=\sum_q\,\sqrt{\frac{\pi}{qL}}\,e^{-aq/2}\,\left(B_q(t)\eta_q+B^{\star}_q(t)\eta^{\dagger}_q\right)
\ee
where
\be
B_q(t)=\frac{1}{\sqrt{K_0}}\left(\cos\eps_q t-i\frac{K_0}{K}\sin\eps_q t\right)\,,
\ee
from which all the local Green's function can be evaluated. We start with the retarded component
\be
G^R_{\varphi\varphi}(t,t')=-i\theta(t-t')\langle\,\left[\varphi(t),\varphi(t')\right]\rangle
\ee
Inserting the time-dependent field expansion~(\ref{eqn:td_field}), we obtain
\bea
G^R_{\varphi\varphi}(t>t')=\frac{2\pi}{L}\,\sum_{q>0}\,\frac{e^{-aq}}{q}\,\mbox{Im}\,B_q(t)\,B_q^*(t')
%=\nonumber\\
%-\frac{2}{K}\,\frac{\pi}{L}\,\sum_{q>0}\,\frac{e^{-aq}}{q}\,\sin\eps_q(t-t')=
%-\frac{2}{K}\,\arctan\left(\Lambda\,(t-t')\right)\nonumber
\eea
Similarly for the Keldysh component
\be
G^K_{\varphi\varphi}(t,t')=-i\langle\left\{\varphi(t),\varphi(t')\right\}
\rangle
\ee
we obtain
\be
 G^K_{\varphi\varphi}(t,t')=-\frac{2\pi\,i}{L}\,\sum_q\,\frac{e^{-aq}}{q}\,\coth\left(\frac{\eps_q}{2T}\right)
\mbox{Re}\,B_q(t)\,B_q^*(t')
\ee
After some algebra and using the fact that
\bea
%\mbox{Re}B_q(t)B_q^*(t')&=& \frac{1}{K_0}
%\left(\cos\eps_q\,t\cos\eps_q t'+\frac{K_0^2}{K^2}\sin\eps_q\,t\sin\eps_q t'\right)\nonumber\\
%\mbox{Im}\,B_q(t)\,B_q^*(t')&=&-\frac{1}{K}\,\sin\eps_q (t-t')
\mbox{Re}B_q(t)B_q^*(t')&=& \frac{1}{K_0}\cos\eps_q\,t\cos\eps_q t'+\nonumber\\
&&+\frac{K_0}{K^2}\sin\eps_q\,t\sin\eps_q t'\\
\mbox{Im}\,B_q(t)\,B_q^*(t')&=&-\frac{1}{K}\,\sin\eps_q (t-t')
\eea
we obtain the following results for the local Green's functions
\bea
G^R_{\varphi\varphi}(t>t')=
-\frac{2}{K}\,\frac{\pi}{L}\,\sum_{q>0}\,\frac{e^{-aq}}{q}\,\sin\eps_q(t-t')=\nonumber\\
=-\frac{2}{K}\,\arctan\left(\Lambda\,(t-t')\right)\nonumber
\eea
and
\begin{widetext}
\bea
G^K_{\varphi\varphi}(t,t')=-\frac{i\pi}{K_0L}\,\sum_q\,\frac{e^{-\alpha q}}{q}\coth(\frac{\eps_q}{2T})
\left[ \cos\eps_q(t-t')\left(1+\frac{K_0^2}{K^2}\right)+
\cos\eps_q(t+t')\left(1-\frac{K_0^2}{K^2}\right)
\right]=\\
=-\frac{i}{K_0}\int \frac{dq}{q}\,e^{-\alpha q}\,\coth(\frac{\eps_q}{2T})
\left[ \cos\eps_q(t-t')\left(1+\frac{K_0^2}{K^2}\right)+
\cos\eps_q(t+t')\left(1-\frac{K_0^2}{K^2}\right)
\right]
\eea
\end{widetext}
The overall result is that the local Green's functions for the boundary field in the OBC case can be obtained from the bulk one by sending
\be
K_0,K\rightarrow \frac{2}{K_0}, \frac{2}{K}
\ee
a result which naturally generalizes the well know equilibrium result. A simple interpretation~\cite{Giamarchi_2003} is that the field in the OBC case corresponds to the $\theta$ field in the translational invariant case (the field $\phi$ being frozen due to the boundary condition~\cite{Giamarchi_2003}), which is known to have correlators with $K\rightarrow 1/K$. On top of this, the field is now living in a semi-infinite space, which explains the extra factor of $2$.
Finally, let us compute the following correlator which will be useful in evaluating the tunneling current
\be
f_{ab}(t_1,t_2)=\langle\,e^{i\gamma\varphi_a(t_1)}\,e^{-i\gamma\varphi_b(t_2)}\rangle
\ee
As before, this can be written only in terms of the local Green's function of the field $\varphi_a(t)$
\be
f(t_1,t_2)_{ab}=\exp\left(-\frac{\gamma^2}{2}\Phi_{ab}(t_1,t_2)\right)
\ee
where we have defined $\Phi_{ab}(t_1,t_2)$ as the following combination of local Green's functions
\bea
\Phi_{ab}(t_1,t_2)&=&
%\left[
\frac{i}{2}G^K_{\varphi\varphi}(t_1,t_1)+\frac{i}{2}G^K_{\varphi\varphi}(t_1,t_1)-iG^K_{\varphi\varphi}(t_1,t_2)
%\right]
+\nonumber\\
& &+i\left[aG^A_{\varphi\varphi}(t_1,t_2)+bG^R_{\varphi\varphi}(t_1,t_2)\right]\nonumber
\eea
Using the above results we can write
\be
f_{ab}(t_1,t_2)=
%\frac{\exp\left(-i\frac{\gamma^2}{2}\left(aG^A(t_1,t_2)+bG^R(t_1,t_2)\right)\right)}
%{\left(1+\Lambda^2\left(t_1-t_2\right)^2\right)^{\tilde{K}^{\rm dual}_{\rm neq}}}
%f(t_1,t_2)
f(t_1,t_2)\frac{\exp \left[-i\Upsilon_{ab}(t_1,t_2)\right]}
{\left(1+\Lambda^2\left(t_1-t_2\right)^2\right)^{\tilde{K}^{\rm dual}_{\rm neq}}}
\ee
where we have defined
\be
\tilde{K}^{\rm dual}_{\rm neq/tr}=\frac{\gamma^2}{4K_0}\left(1\pm\frac{K_0^2}{K^2}\right)
\ee
as well as
\bea
f(t_1,t_2)=\left[
\frac{\left(1+4\Lambda^2 t_1^2\right)\left(1+4\Lambda^2 t_2^2\right)}
{\left(1+\Lambda^2(t_1+t_2)^2\right)^2}\right]^{\tilde{K}^{\rm dual}_{\rm tr}/2}\\
\Upsilon_{ab}(t_1,t_2)=\frac{\gamma^2}{2}\left(aG_{{\varphi\varphi}}^A(t_1,t_2)+bG_{\varphi\varphi}^R(t_1,t_2)\right)
\eea

%\bibliography{mybiblio}

%

\end{document}